\input harvmac

\Title{\vbox{\baselineskip12pt
\hbox{YCTP-P21-97}
\hbox{hep-th/9710230}
}}
{\vbox{\centerline{ $K$-theory and Ramond-Ramond  charge}}}

\centerline{Ruben Minasian and Gregory Moore}

\bigskip
{\vbox{\centerline{\sl Department of Physics, Yale University}
\vskip2pt
\centerline{\sl New Haven, CT 06520}}
\centerline{ \it ruben.minasian@yale.edu }
\centerline{ \it gregory.moore@yale.edu }

\bigskip
We discuss the relation between the Ramond-Ramond
charges of D-branes and the topology of
Chan-Paton vector bundles. We show that a topologically
nontrivial normal bundle induces RR charge and that
the result fits in perfectly with the proposal that D-brane
charge is the topology of the Chan-Paton bundle, regarded as an
element of
$K$-theory.

\vskip .3in
\Date{\vbox{\baselineskip12pt
\hbox{October 29, 1997}}}

%
\lref\polchinski{J. Polchinski, ``Dirichlet branes and Ramond-Ramond
charges,'' Phys. Rev. Lett. {\bf 75} (1995) 4724, hep-th/9510017.}
\lref\bvs{M. Bershadsky, V. Sadov, and C. Vafa  ``D-branes and
topological field theories,'' Nucl. Phys. {\bf B463} (1996) 420,
hep-th/9511222.}
\lref\ghm{M. Green, J. Harvey, and G. Moore,
``I-Brane inflow and anomalous couplings on D-branes,''
Class. Quant. Grav. {\bf 14}(1997) 47,  hep-th/9605033.}
\lref\GrHa{P.~ Griffiths and J.~ Harris, {\it Principles of
Algebraic Geometry}, ch. 3, J.Wiley and Sons, 1978. }
\lref\hirz{F. Hirzerbruch, {\it Topological Methods in Algebraic
Geometry},
Springer, 1978.}
\lref\sping{H.B. Lawson and M.L. Michelsohn, {\it Spin Geometry},
Princeton
Univ. Press, 1989.}
\lref\karoubi{M. Karoubi, {\it $K$-theory}, Springer, 1978.}
\lref\morrison{D. Morrison, ``The Geometry Underlying Mirror
Symmetry,''
alg-geom/9608006; Proc. European Algebraic Geometry Conference
(Warwick,
1996).}
\lref\hmii{J. Harvey and G. Moore, ``On the algebra of BPS states,''
hep-th/9609017.}
\lref\bbs{K. Becker, M. Becker and A. Strominger, ``Fivebranes,
Membranes and
Non-Perturbative String Theory,''  Nucl. Phys. {\bf B456} (1995) 130,
hep-th/9507158.}
\lref\yin{Y.K. Cheung and Z. Yin, ``Anomalies, Branes, and
Currents,''
hep-th/9710206.}
\lref\wittfive{E. Witten, ``Five-Brane Effective Action In
M-Theory,''
hep-th/9610234.}
\lref\lw{W. Lerche and N. Warner, ``Index Theorems in $N=2$
Superconformal
Theories,'' Phys. Lett. {\bf B205} (1988) 471.}
\lref\hair{See, S. Coleman,  J. Preskill and F. Wilczek,
``Quantum Hair on Black Holes,''  Nucl. Phys. {\bf B378} (1992) 175,
hep-th/9201059, and references therein. }
\lref\douglas{M. Douglas, ``Branes within branes,'' hep-th/9512077.}
\lref\conifold{A. Strominger, ``Massless black holes and conifolds
in string theory,''  Nucl. Phys. {\bf B451}(1995) 96,
hep-th/9504090.}
\lref\wwwf{D.~S.~Freed and E.~Witten, ``Anomalies in string theory
with D-branes,'' hep-th/9907189\semi E.~Witten, ``Baryons and
branes in anti de Sitter space,'' JHEP {\bf 9807}, 006 (1998)
hep-th/9805112\semi E.~Witten, ``D-branes and K-theory,'' JHEP {\bf
9812}, 019 (1998) hep-th/9810188.}
%

\newcount\figno
\figno=0
\def\fig#1#2#3{
\par\begingroup\parindent=0pt\leftskip=1cm\rightskip=1cm\parindent=0pt

\baselineskip=11pt
\Gammalobal\advance\figno by 1
\midinsert
\epsfxsize=#3
\centerline{\epsfbox{#2}}
\vskip 12pt
{\bf Fig. \the\figno:} #1\par
\endinsert\endgroup\par
}
\def\figlabel#1{\xdef#1{\the\figno}}
\def\encadremath#1{\vbox{\hrule\hbox{\vrule\kern8pt\vbox{\kern8pt
\hbox{$\displaystyle #1$}\kern8pt}
\kern8pt\vrule}\hrule}}

\overfullrule=0pt
%


\def\inbar{\,\vrule height1.5ex width.4pt depth0pt}
\def\IB{\relax{\rm I\kern-.18em B}}
\def\IC{\relax\hbox{$\inbar\kern-.3em{\rm C}$}}
\def\ID{\relax{\rm I\kern-.18em D}}
\def\IE{\relax{\rm I\kern-.18em E}}
\def\IF{\relax{\rm I\kern-.18em F}}
\def\IG{\relax\hbox{$\inbar\kern-.3em{\rm G}$}}
\def\IH{\relax{\rm I\kern-.18em H}}
\def\II{\relax{\rm I\kern-.18em I}}
\def\IK{\relax{\rm I\kern-.18em K}}
\def\IL{\relax{\rm I\kern-.18em L}}
\def\IM{\relax{\rm I\kern-.18em M}}
\def\IN{\relax{\rm I\kern-.18em N}}
\def\IO{\relax\hbox{$\inbar\kern-.3em{\rm O}$}}
\def\IP{\relax{\rm I\kern-.18em P}}
\def\IQ{\relax\hbox{$\inbar\kern-.3em{\rm Q}$}}
\def\IR{\relax{\rm I\kern-.18em R}}
\font\cmss=cmss10 \font\cmsss=cmss10 at 7pt
\def\IZ{\relax\ifmmode\mathchoice
{\hbox{\cmss Z\kern-.4em Z}}{\hbox{\cmss Z\kern-.4em Z}}
{\lower.9pt\hbox{\cmsss Z\kern-.4em Z}}
{\lower1.2pt\hbox{\cmsss Z\kern-.4em Z}}\else{\cmss Z\kern-.4em
Z}\fi}
\def\IGa{\relax\hbox{${\rm I}\kern-.18em\Gamma$}}
\def\IPi{\relax\hbox{${\rm I}\kern-.18em\Pi$}}
\def\ITh{\relax\hbox{$\inbar\kern-.3em\Theta$}}
\def\IOm{\relax\hbox{$\inbar\kern-3.00pt\Omega$}}

\font\zfont = cmss10 
\font\litfont = cmr6

\def\bigone{\hbox{1\kern -.23em {\rm l}}}
\def\ZZ{\hbox{\zfont Z\kern-.4emZ}}
\def\DS {D\!\!\!\!/}
\def\half{{\litfont {1 \over 2}}}

\def\CW{{\cal W}}
\def\CS{{\cal S}}
\def\CN{{\cal N}}

\def\ra{\rightarrow}

\def\ch{{\rm ch}}
\def\IR{\relax{\rm I\kern-.18em R}}
\font\cmss=cmss10 \font\cmsss=cmss10 at 7pt

\newsec{Introduction}

When string theory is compactified on a spacetime
$\CS$ the low energy effective theory typically
contains many $U(1)$ gauge fields. In this paper we
will focus on the RR gauge fields arising from KK
reduction of the RR gauge fields of  ten-dimensional
type IIA and IIB supergravity.
The vector space of these gauge fields
is dual to a space of harmonic forms in $\CS$, and
hence RR charges take values in $H^*(\CS;\IR)$.
Quantization of RR charge implies that, in fact,
the charge takes values in the $\IZ$-module
$H^*(\CS; \IZ)$.\foot{Actually, it appears that one should
replace $\IZ \rightarrow {1 \over  N} \IZ$
for some sufficiently large $N$. }

Perturbative string states
all have RR charge zero, but in nonperturbative
string theory part of the spectrum is
associated with states formed by wrapping D-branes
around supersymmetric cycles
$\CW$ in $\CS$ \refs{\conifold, \bbs}.
The data specifying a wrapped D-brane state includes
a choice of vector-bundle $E \rightarrow \CW$,
called the Chan-Paton bundle, and a connection on
$E$. In contrast to perturbative states,
wrapped  D-branes are charged under the RR gauge fields
\polchinski.  In this note we will derive a formula for
the RR charge of the wrapped D-brane in terms of the
topology of the embedded cycle
$f: \CW\hookrightarrow \CS $
and  the topology of $E$.
Our main result is
\eqn\newcharge{Q = {\rm ch} \left( f_{!}E  \right) \sqrt{{\widehat
A}({\cal
TS})}  }
where $\CT\CS$ is the tangent bundle to
spacetime and $f_{!}$ is the $K$-theoretic
Gysin map. As we discuss below this
strongly suggests that the proper conceptual
home for RR charge is in fact the $\IZ$-module
$K(\CS)$ rather than $H^*(\CS;{1 \over  N} \IZ)$.
The result \newcharge\ generalizes the result
of \ghm. The main difference is that in \ghm\ the normal
bundles to the branes were assumed to be topologically
trivial.

A closely related paper with
some overlapping results has recently appeared  \yin.

\bigskip
\noindent
{\it Note added for version 3.}  Unfortunately, the note added for
version 2 of this paper has caused a certain amount of confusion.
In view of recent progress on the relation of D-branes to K-theory
we would like to add a further clarification.  The results of this
paper, while similar to those of \yin, differ in one important
detail, namely, in  the presence of the factor $e^{\half d}$ in
equations $(2.4), (2.7), (3.1),(3.3)$, etc. While our original
explanation for this factor (based on a misidentification of the
quantum numbers of the $I$-brane fermions under the normal bundle
structure group) was wrong,  in fact equations  $(2.4), (2.7),
(3.1),(3.3)$, and,  most importantly equation $(1.1)$ indeed turn
out to be correct. A proper understanding of the factor $e^{\half
d}$ requires noting that the normal bundle to the brane worldvolume
must carry a   $Spin^c$ structure and also requires introducing the
NS $B$-field.  This has been lucidly explained in
\wwwf.

\newsec{Derivation}

The derivation of \newcharge\  proceeds along
the lines of \ghm, except we will relax the assumption
made in \ghm\ that the normal bundles have
trivial topology.
In general, if $f: \,\,\, X \hookrightarrow Y$
is an embedding,  we  will
denote by $\CN(X,Y)$
the normal bundle defined by:
\eqn\nrmlbdle{
0 \ra TX \  {\buildrel {f_{\ast}\!\!\! }
 \over \longrightarrow} \ TY \ra \CN(X,Y) \ra 0.}

Let $f: \CW \hookrightarrow \CS$ be the
embedding of the D$p$-brane worldvolume into
spacetime.
\foot{We assume embeddings for convenience. The
discussion should be generalized to allow immersions.}
The (anomalous) coupling on the
worldvolume  of $N$ coincident D-branes may be
written in the form:
\eqn\inflow{I_{\cal W}=\int_{\CW} c \wedge Y(\CF, g), }
where $\CW$ is the $p+1$-dimensional
world-volume of the  brane,
$c=f^{\ast}C$ is the pullback of the total RR potential $C$,
$\CF = F - f^*B$ where $B$ is the NS 2-form potential,
$F$ is the Hermitian
field strength of the $U(N)$ gauge field on the
brane, and $g$ is the restriction of the spacetime metric
to the brane. Defining ${\rm ch}( E) =
{\rm tr}_N\exp\left( {\CF \over 2\pi}\right)$,
\ghm\ obtained the formula:
\eqn\charge{Y(\CF,g) = {\rm ch}( E) f^* \sqrt{{\widehat A}(\CT\CS)}.
}

As in \ghm\ we will deduce the anomalous coupling
$Y(\CF, g)$ from an anomaly inflow argument.
Consider two branes intersecting on $\CI \equiv \CW_1 \cap \CW_2$
such that
there are chiral fermions on $\CI$.
To fix ideas, consider two sevenbranes intersecting on a
fivebrane.\foot{In
the IIB theory, $\CN(\CW , \CS)$ is always even.
In the IIA theory it is odd. We derive formulae for the charges in
IIB and
relate
them to IIA by T-duality along some circle $S^1$.}
In this situation there are massless chiral fermions
moving on $\CI$. Let us consider the quantum numbers
of these fermions.
First, using standard D-brane techniques one finds that
if the Chan-Paton bundles on $\CW_1, \CW_2$
have ranks $N_1, N_2$, respectively, the fermions are in
a hypermultiplet of the gauge group $U(N_1) \times U(N_2)$
in the representation $(N_1, \bar N_2) \oplus (\bar N_1, N_2)$.

There is another set of fermion
quantum numbers associated with the normal bundles
$\CN_i \equiv  \CN(\CI, \CW_i)$
of $\CI$.  According to \bvs\  the scalar fields in the
hypermultiplets are sections of these normal bundles.
Let us see what this implies for the quantum numbers
of the fermions.

We choose local coordinates at the intersection so that
the two sevenbranes are in the $67$ and $89$ directions.
 The little group of the $5+1$ theory includes
$G=Spin(4)_{2345} \times \bigl[
Spin(2)_{67} \times Spin(2)_{89}\bigr] $.
The latter factor is a subgroup of
$Spin(4)_{6789}$. We decompose the Lie algebra as:
$so(4)_{6789} = su(2)_- \oplus su(2)_+$
where    $su(2)_-$ is
the $\CR$-symmetry of $d=6, \CN=(1,0)$ supersymmetry.
The presence of the $\CI$ brane breaks the internal
symmetry to $Spin(2)_{67} \times Spin(2)_{89}$.
Denote the generators of
$so(2)_- \oplus so(2)_+$ as
$T_-=\half(T_{67}-T_{89})$ and
$T_+=\half(T_{67} + T_{89})$, respectively.
Globally, $Spin(2)_{67}$ and $Spin(2)_{89}$ are the
structure groups of principal $SO(2)$ bundles over
$\CI$. The normal bundles $\CN$ are associated by
the vector representation.

Now consider the string oscillator quantization.
The 5+1 worldvolume bosons on the
$\CI$-brane come from the NS sector.
Quantization of the  fermion zeromodes
$\psi_0^{6,7,8,9}$
 gives   two complex bosons
transforming under $G$ as:
$(1,1;T_+ + \half, T_+ -\half ) + (1,1;T_+ - \half, T_+ +\half )$.
The 5+1 worldvolume fermions on the
$\CI$-brane come from the Ramond sector.
Quantization of
the fermions zeromodes
$\psi_0^{2,3,4,5}$ gives states
 transforming under $G$
as $(1,2;T_+,T_+)$.
In order for the bosons to be sections of the
normal bundle the hypermultiplet must have
charge $1/2$ under $T_+$.
We conclude that the worldvolume
fermions are   sections of
the spinor bundles associated to $\CN_i$.
\foot{This kind of reasoning
 resolves a small paradox in $F$-theory.
In $F$-theory compactifications,
it can happen that a supersymmetric cycle
$\CW$, or even spacetime, does not admit a spin
structure. As an example,
consider $F$-theory compactification to six dimensions
where the base of the elliptically fibered CY 3-fold is
$\IP^2$ blown up at an even number of points. However,
the sevenbranes induce a $Spin^c$ structure allowing the
presence of the fermions. Similarly, the fermions of the $M$-theory
fivebrane
are   valued
in the spinor bundle associated to the normal bundle \wittfive.}

Now that we have the quantum numbers of the
chiral fermions on $\CI$ we can compute the anomaly
applying the standard descent procedure to:\foot{Here
we correct a factor of two in \ghm. This is discussed at length in
\yin.}
\eqn\inters{X({\cal I}) = {\rm ch}_{N_1}( E_1) {\rm ch}_{N_2}( E_2)
e^{\half d_1 + \half d_2} {\widehat A}({\cal TI}).}
where $d_i = c_1(\CN_i)$.

We now rewrite \inters\ in a form which makes
clear how to cancel the $\CI$-brane fermion anomaly
via inflow due to the anomalous couplings \inflow\
on $\CW_i$.  When the intersections of $\CW_1$ and $\CW_2$ are
chiral,   the
brane configurations fill all spacetime dimensions. It follows that
\eqn\seqq{0 \ra {\cal TI} \ra {\cal TW}_1\oplus {\cal TW}_2
{\buildrel {\psi} \over \longrightarrow} {\cal TS}\ra 0}
is an exact sequence where $\psi(\xi_1 \oplus \xi_2)=\xi_1 - \xi_2.$
Recalling that ${\widehat A}$ is a multiplicative
characteristic class, ${\widehat A}(E \oplus F)={\widehat A(E)} \cdot
{\widehat
A(F)}$, it
is easy to see that
\eqn\ahati{{\widehat A}({\cal TI}) = f^* { {\widehat A}({\cal TW}_1)
\cdot
{\widehat A}({\cal TW}_2)
\over {\widehat A}({\cal TS})}  }
and  hence the anomaly from
\inters\  can be cancelled by modifying\foot{An alternative proof
invloves
the isomorphisms of the normal bundles $\CN({\cal I}, {\cal W}_1)
\cong
{\cal N} ({\cal W}_2, {\cal S}) \vert_{\CI} $ and as a consequence,
$\CN(\CI,  \CS) \cong \CN( \CW_1, \CS)\vert_{\CI}
\oplus \CN(\CW_2, \CS)\vert_{\CI}.$}
\eqn\modify{
Y \ra Y' =  Y e^{\half d}
{{\widehat A}({\cal TW}) \over
f^* {\widehat A}({\cal TS})}
}
in \charge. Here
$d$ is a degree two class defining a $Spin^c$ structure on $\CW$. It
can induce
charge shifts of degree two on
D-brane worldvolumes along the lines of  \refs{\bvs, \douglas, \ghm}.

Note that this correction is just $\left( {\widehat A}(\CN(\CW, \CS))
\right)^{-1}$ and hence reflects the contribution from the normal
bundle.One
might wonder why we should worry about the full sequence $  {\widehat
A}\left(
\CN \right).$ The answer is, as remarked in \wittfive, that one
should consider
families of branes.

\newsec{Interpretation of the result}

We would now like to explore the conceptual
meaning of the modification \modify.
We write  the modified coupling on the brane in the form
\eqn\infnew{I_{\cal W}=\int_{\CW} c
\wedge Y^{\prime}({\cal W})
= \int_{\CW} c \wedge {\rm ch}( E)
 {{\widehat A}({\cal TW}) e^{{1 \over 2}d}
\cdot  {1 \over  f^{\ast}\sqrt{{\widehat A}({\cal TS})}}}
}
To obtain the D-brane charge, one studies
the RR equation of motion and Bianchi identity
coming from \infnew. Note that the naive equation of motion:
\eqn\eomi{
d*F(C) = \delta({\cal W}
\hookrightarrow {\cal S}) Y^{\prime}({\cal W})
}
 is valid only when the normal
bundle is trivial. In general, the LHS of \eomi\    is defined
on $ {\cal S}$, while the RHS involves only quantities defined on  $
{\cal W}$.
The correct equation is:
\eqn\eomc{d*F(C) = f_{\ast}\left(
{\rm ch}( E) {\widehat A}({\cal TW}) e^{{1 \over 2}d} \cdot
{1 \over  f^{\ast}\sqrt{{\widehat A}({\cal TS})} } \right).
}
Here we have introduced the Gysin map, or
 push-{\it forward}  $f_{\ast}$ acting on {\it
cohomology}.\foot{Strictly
speaking, we are dealing not with
cohomology classes but with ``currents''  \GrHa.
We need a $\delta$-function representative of the
RHS of \eomc.  For further
discussion see \yin.}
The map $f_*$ is defined  using   the Poincar\'{e} duality map
${\cal D}$ twice - first  on  ${\cal W}$, then on ${\cal S}$. That is
  $f_{\ast} = {\cal D}_{\cal S} ^{-1} f_*^{h} {\cal D}_{\cal W}$
where on the RHS $f_*^{h}$ is the natural push-forward on
{\it homology}. Thus:
\eqn\flows{f_{\ast}: H^k( {\cal W},  \ZZ)  \rightarrow H^{k+r}( {\cal
S},
\ZZ),}
where $r= {\rm rank} {\cal N}(\CW,\CS)$.   The new equation
of motion for the RR potential \eomc\ is consistently defined in the
bulk.  It
is easy to check that  the RHS of  \flows\ has the right degree.

We can proceed to write the formula for the pairing of the charge
vector
$Q \in H^{\ast} ({\cal S})$ with a homology cycle   $\Gamma \in
H_*(\CS)$
\eqn\prodd{(Q, \Gamma) = \int_{ \Gamma} f_{\ast}\left(  {\rm ch}( E)
{\widehat
A}({\cal TW}) e^{{1 \over 2}d} \cdot  {1 \over
f^{\ast}\sqrt{{\widehat
A}({\cal TS})}} \right).}
To avoid cluttering notation we omit the pullback to
$\Gamma$.
To evaluate this integral, we   note that     for $ \phi \in H^*(
{\cal W},  \ZZ) $ and $ \theta \in H^*( {\cal S},  \ZZ) $:
\eqn\useful{f_{\ast}\left( \phi \wedge f^{\ast}  \theta \right) =
f_{\ast} \phi
\wedge \theta.}
This follows from  the Thom isomorphism:
\eqn\fff{\eqalign{\phi \in H^*( {\cal W},  \ZZ) : \,\,\,\,\,
&f^{\ast}f_{\ast}\phi = \chi
\wedge \phi\cr
\theta \in H^{* }(\CS ,  \ZZ) : \,\,\,\,\, &
f_{\ast}f^{\ast}\theta = \eta \wedge \theta \cr }}
where $\chi$ is the Euler character of the normal bundle and $\eta$
is the
Poincar\'{e} dual of the zero section. Thus \prodd\ becomes
\eqn\prodc{(Q, \Gamma) = \int_{ \Gamma} f_{\ast}\left(  {\rm ch}( E)
{\widehat
A}({\cal TW}) e^{{1 \over 2}d} \right)   {1 \over  \sqrt{{\widehat
A}({\cal
TS})}} .}

The next step involves the use of a version of  the
Grothendieck-Riemann-Roch theorem for compact oriented differentiable
manifolds.
We work in the IIB theory so that $\CN(\CW,\CS)$ can
be considered as a complex vector bundle on $\CW$.
It inherits an Hermitian structure from $g$.
Under these circumstances we can apply a theorem of
Atiyah and Hirzebruch \refs{\hirz, \sping, \karoubi}
which states that for a certain class $ f_{!}E$ of  $K(\CS)$
(defined below) we have
\eqn\push{
f_{\ast}\left( {\rm ch}(E) {\widehat A}({\cal TW}) e^{{1\over 2}d}
\right)=
{\rm ch} \left( f_{!}E \right) {\widehat A}({\cal TS}) .}
Here $d$ is an element in $H^2({\cal W}, \ZZ)$ whose reduction mod2
is
$w_2({\cal W}) - f ^*w_2({\cal S}) = w_2(\CN)$.  In the special  case
of
connected almost
complex manifolds,
one gets a very nice expression
$d = c_1({\cal W}) - f ^*c_1({\cal S})=-c_1(\CN).$
\foot{This can probably also be deduced using
the results in  \lw, where
the role of  ${1 \over 2}c_1(\CS)$ in $\CN=2$
string compactification was investigated.  }
The map $f_!$ is an isomorphism analogous to the Thom isomorphism for
cohomology. Denoting the pull-back operation for bundles
with superscript $!$   and
specializing to our case, we can define $f_!$ as
\eqn\low{
f_! E  = \pi^! E  \otimes \delta({\cal N}).
}
Here $\pi: \,\,\, \CN \ra \CW$ is the projection,
so $\pi^!E$ can be extended to a bundle on $\CS$.
$\delta(\CN)$ is a $K$-theoretic analogue of the Thom
class, defined as follows. First,
$\delta({\cal N})$ is an element of  relative $K$ theory
for bundles on
the tubular neighborhood of $\CW$ relative to the
sphere bundle, again extended to $K(\CS)$.\foot{$f_!$ defines
an  isomorphism map
$f_!: \,\,\, K(\CW) \ra K_{\rm cpt}(\CS)$  \sping.
A more physical discussion should replace
``relative $K$-theory'' by some version of
``relative $K$-theory with rapid decrease in the vertical
direction''. } In
particular,
$\delta({\cal N})$ is defined as the isomorphism
class of the  triple:
\eqn\thom{\eqalign{\delta({\cal N})
&\cong (-1)^{r}\left[  \Lambda^{\rm even}_{\IC} \pi^*\CN\
 {\buildrel {\sigma(s)} \over \longrightarrow}\
\Lambda^{\rm odd}_{\IC} \pi^*\CN \right],}}
where  $s$ is a vector in the unit sphere bundle,
and, using the Hermitian structure on $\CN$ we can
define a contraction $\iota(s)$ and hence the isomorphism:
$\sigma(s)= s\wedge - \iota(s)$. If $\CN$ admits
a spin structure this is Clifford multiplication by
$s$. For further
details and definitions see \refs{\hirz,\sping, \karoubi}.

Using \push, we finally arrive at our main result:
As an element of $H^*(\CS)$ the RR charge
associated to a D-brane wrapping a supersymmetric
cycle embedded in spacetime by
$f: \CW \hookrightarrow \CS$ with Chan-Paton
bundle $E \rightarrow \CW$ is given by:
\eqn\newchargea{Q = {\rm ch} \left( f_{!}E \right) \sqrt{{\widehat
A}({\cal
TS})}.}

This result  has a natural $K$-theoretic interpretation,
first remarked (in the context of the result of
\ghm) independently by Graeme Segal and Maxim Kontsevich.
It is well-known that the Chern character is a ring isomorphism
from $K(X, \IQ )$ to the even-dimensional cohomology
$H^{\rm even}(X;\IQ)$.
Both $K(X, \IQ )$ and $H^{\rm even}(X;\IQ)$
have natural bilinear pairings. The pairing on
$H^{\rm even}(X;\IQ)$ is just given by
$(\omega_1, \omega_2)_{DR} = \int_{X} \omega_1 \wedge \omega_2$,
while the pairing on $K(X)$ is given by the
index of the Dirac operator:
$(E_1, E_2)_K \equiv {\rm ind}{\DS}_{E_1 \otimes E_2}$.
This may be written in terms of the DeRham pairing as:
\eqn\bilinear{(E_1, E_2)_K
= \left ({\rm ch} \left( E_1 \right)\sqrt{{\widehat A}(TX)} , \,\,
{\rm ch} \left( E_2\right) \sqrt{{\widehat A}(TX)}  \right)_{DR}.
}
That is, the modified   Chern isomorphism
\eqn\modchern{
E \rightarrow \ch(E) \sqrt{ \hat{A}(TX) }
}
is an {\it isometry} with respect to the natural bilinear
pairings on $K(X)$ and $H^*(X)$.
Thus, the most natural statement of
the result \newchargea\ is that the D-brane charge is
given by $f_!(E) \in K(\CS)$.

\newsec{Discussion and conclusion}

We conclude by making a few remarks and pointing to
a few possible directions for further research. First, we find the
identification of RR charge with classes
in $K$-theory philosophically quite natural and harmonious.
It is perfectly consistent with the interpretation
\refs{\morrison, \hmii}
of Chan-Paton bundles as coherent sheaves, in the complex
setting. However, the $K$-theoretic interpretation
is more general since the derivation does not
assume any complex structure on $\CW$ or $\CS$.

Nevertheless, it would be nice to understand the
appearance of $f_!(E)$ as defined in
\low\  (as well as the coherent sheaves of  \refs{\morrison, \hmii})
in a more physical language.  We hope to
return to this in a future publication.

Strictly speaking, our argument for
the identification of
$Q$ with $f_!(E)$ only applies over
$\IQ$. However, both $H^*(X;\IZ)$ and
$K(X)$ can have torsion, and the above
discussion emphasizes the possibility that
D-branes can have torsion charges
(e.g. $\IZ_N$ charges) and raises the question as to
which group has the physically
correct torsion classes. There are cases
where the torsion in $K(X)$ differs from
the torsion in $H^*(X)$. The most elementary
example is $X=S^1$, where
$\widetilde{KO}(X) =\IZ_2$
while the cohomology has no torsion.
As an example of a  D-brane with nonzero torsion
charge one could consider a type I wrapped
D1 string whose Chan-Paton bundle is
the Mobius bundle. If this twisting has
physical consequences (inducing, say, nontrivial
Aharonov-Bohm phases in scattering) then
the torsion of $K$-theory is the relevant one.
 It would be very interesting
to study these torsion charges more generally;
they would be the D-brane analogues of the
 the $\IZ_N$ charges on black holes and solitons that were
explored in \hair.

Another, more practical, application of
\newchargea\ might be to anomaly cancellation
in various backgrounds in $M$- and $F$-theory.
It is possible that
the modifications of D-brane charges due to
normal bundle topology will shed light on some
of the difficulties with fivebrane anomalies that
have been discussed in \wittfive.

\bigskip
\centerline{\bf Acknowledgements}\nobreak
\bigskip

We would like to thank M. Kontsevich and G. Segal
for the essential remark mentioned above.
We also thank  A. Gerasimov, M. Green, J. Harvey,  W. Lerche,
P. Mayr, C. Schweigert, R. Szczarba, S. Shatashvili, S. Theisen and
E. Witten
for discussions. GM would like to thank
CERN  and the Aspen Center for Physics
for hospitality during part of this
work. This work  is supported by
DOE grant DE-FG02-92ER40704.

\listrefs
\end